\documentclass[runningheads]{llncs}
\newcommand{\MUSICNTWRK}{ {\texttt{MUSICNTWRK}}}
\newcommand{\musicntwrk}{ {\texttt{musicntwrk}}}
\newcommand{\data}{ {\texttt{data}}}
\newcommand{\networks}{ {\texttt{networks}}}
\newcommand{\harmony}{ {\texttt{harmony}}}
\newcommand{\timbre}{ {\texttt{timbre}}}
\newcommand{\mlutils}{ {\texttt{ml\_utils}}}
\newcommand{\utils}{ {\texttt{utils}}}
\newcommand{\plotting}{ {\texttt{plotting}}}

\usepackage{graphicx}

\usepackage[dvipsnames]{xcolor}
\usepackage[htt]{hyphenat}

\usepackage[T1]{fontenc}

\definecolor{Mygreen}{HTML}{73C000}
\definecolor{Mygrey}{HTML}{4C463E}
\definecolor{Mypurple}{HTML}{DF89FF}
\definecolor{Mycyan}{HTML}{00C4FF}
\definecolor{Mypink}{HTML}{FF5584}
\definecolor{Myorange}{HTML}{FF8805}

\begin{document}
\title{\MUSICNTWRK: data tools for music theory, analysis and composition}
\titlerunning{\MUSICNTWRK}
\author{Marco Buongiorno Nardelli\inst{1,2,3,4,5}\orcidID{0000-0003-0793-5055}}
\authorrunning{M. Buongiorno Nardelli}
\institute{CEMI, Center for Experimental Music and Intermedia, University of North Texas, Denton, TX 76203, USA \and
iARTA, Initiative for Advanced Research in Technology and the Arts, University of North Texas, Denton, TX 76203, USA
\and
Department of Physics, College of Science, University of North Texas, Denton, TX 76203, USA
\and
Division of Composition Studies, College of Music, University of North Texas, Denton, TX 76203, USA
\and
IM\'eRA - Institut d'\'Etudes Avanc\'ee of Aix-Marseille Universit\'e, Marseille 13004, France \and
Laboratoire CNRS-PRISM, 13402 Marseille, France\\
\email{mbn@unt.edu}\\
\url{http://www.musicntwrk.com, http://www.materialssoundmusic.com}}
\maketitle              % typeset the header of the contribution
\begin{abstract}
We present the API for \MUSICNTWRK, a python library for pitch class set and rhythmic sequences classification and manipulation, the generation of networks in generalized music and sound spaces, deep learning algorithms for timbre recognition, and the sonification of arbitrary data. The software is freely available under GPL 3.0 and can be downloaded at www.musicntwrk.com or installed as a PyPi project ({\tt pip install musicntwrk}).

\end{abstract}
\linespread{0.98}
\section{Introduction}
Big data tools have become pervasive in virtually every aspects of culture and society. In music, application of such techniques in Music Information Retrieval applications are common and well documented. However, a full approach to musical analysis and composition is not yet available for the music community and there is a need for providing a more general education on the potential, and the limitations, of such approaches. From a more fundamental point of view, the abstraction of musical structures (notes, melodies, chords, harmonic or rhythmic progressions, timbre, etc.) as mathematical objects in a geometrical space is one of the great accomplishments of contemporary music theory. Building on this foundation, we have generalized the concept of musical spaces as networks and derive functional principles of compositional design by the direct analysis of the network topology. This approach provides a novel framework for the analysis and quantification of similarity of musical objects and structures, and suggests a way to relate such measures to the human perception of different musical entities. 
The original contribution of this work is in the introduction of the representation of musical spaces as large-scale statistical mechanics networks: uncovering their topological structure is a fundamental step to understand their underlying organizing principles, and to unveil how classifications or rule-based frameworks (such as common-practice harmony, for instance) can be interpreted as emerging phenomena in a complex network system. Results from this research, the theoretical and technical foundation for this paper and some application of this techniques in composition and analysis can be found in Ref. ~\cite{ref_article1,ref_article2,ref_article3}. In the following we give the full API of the software library after summarizing some important definitions.

\section{Background}\label{background}
Network analysis methods exploit the use of graphs or networks as convenient tools for modeling relations in large data sets. If the elements of a data set are thought of as nodes, then the emergence of pairwise relations between them, edges, yields a network representation of the underlying set. Similarly to social networks, biological networks and other well-known real-world complex networks, entire data-set of sound structures can be treated as a network, where each individual descriptor is represented by a node, and a pair of nodes is connected by a link if the respective two objects exhibit a certain level of similarity according to a specified quantitative metric. 
Pairwise similarity relations between nodes are thus defined through the introduction of a measure of  distance   in the network: a  metric.
In \MUSICNTWRK\ we use the Euclidean norm (generalized Pythagoras theorem in N-dimensions) to quantify similarity between descriptors:
\begin{equation}
    \label{norm}
    \textrm{distance}(I,J) = \sqrt{\sum_i \left (x^I_i-x^J_i\right )^2},
\end{equation}
where ${\bf x}$ is the chosen sound descriptor for sound $I$ and $J$.
We specialize in the following the metric for the two major musical spaces in \MUSICNTWRK\ and define two of the operators that are central to many of the algorithms in the \MUSICNTWRK\ package:

\begin{enumerate}
\item
Metric in the interval vectors space: the distance operator for interval vectors can be written as:

$$d({\bf x},{\bf y}) = \sqrt{\sum_i{({\bf x}_i-{\bf y}_i)^2}}$$

where {\bf x} and {\bf y} are interval vectors of dimension INT(N$_C$/2), where N$_C$ is the cardinality of the pcs (number of total pitches). This is a quantity that measures the change in the harmonic content of two pcs, and thus contains a quantification of the rules of harmony in arbitrary musical spaces.
\item
Metric in the ordered pcs space (voice leading): the distance operator for pcs provides instead a quantification of voice leading, the study of the linear progression of individual melodic lines at the foundation of counterpoint. For this we use minimal Euclidean voice leading distance [1] for arbitrary TET-notes temperaments (12, 24, etc.):

$$d_{\textrm{min}}({\bf x},{\bf y}) = \textrm{min}_{\textrm{\bf TET}_j}\sqrt{\sum_i{({\bf x}_i-({\bf y}_j\pm \textrm{\bf TET}_j))^2}}$$

Here {\bf x} and {\bf y}  are the pcs in normal order and TET=(0,0,$\dots$,0,$\pm$TET,0,$\dots$),  a vector of dimension N$_C$ that raises or lowers the $j^\textrm{th}$ pitch of the ordered pcs by TET. It is easy to verify that this definition of distance operator is equivalent to finding the minimal distance between all possible cyclic permutations of the pcs. Such definition is easily extended to non-bijective voice-leadings, by an iterative duplication of pitches in the smaller cardinality pcs, and then looking for the multiset that produces the minimal distance. Note that, although we use eucledean distance throughout this discussion, the metric can be chosen by the user from the full palette of metrics available in {\tt sklearn} through the {\tt distance} variable.
\end{enumerate}
From the definition of metrics above we can introduce two essential operators in the vector spaces of interval vectors or ordered pcs:

\begin{enumerate}
\item
The distance operators: O(\{$n_i$\}) as defined in \MUSICNTWRK\ is an operator that raises or lowers by an integer $n$ the $i^\textrm{th}$ component of a vector. In the interval vector space, these are vector operators of dimension INT(N$_C$/2); in the voice-leading space, they have dimension N$_C$.  With this definition, if {\bf x} is transformed into {\bf y} by O(\{$n_i$\}), then:

$$d({\bf x},{\bf y}) = \sqrt{\sum_i n_i^2} $$

The distance operator defined above is assumed to be raising or lowering by the specified amount one of the pitches of the pcs (or components of interval vector (IV)) with no information on the positional ordering (i is left unspecified). As such it is a function that acts upon the vector of all the permutations of the input pcs. For instance, the voice leading operator, O(1) applied to the [0,4,7] pcs (C Maj chord) generates the following chords: [0, 3, 7] (C min), [0, 4, 6] (C incomplete half-diminished seventh), [0, 4, 8] (C augmented), [5, 7, 0] (F quartal trichord), [1, 4, 7] (C\# diminished), [4, 7, 11] (E min). If the number of specified components is smaller than the cardinality of the pcs or the dimension of the IV, all unspecified components are assumed to be 0.  
\item
Based on the metric in voice leading space, we can introduce the normal-ordered voice leading operator VL({\bf n}), that given a normal-ordered pcs, transform it into the successive normal-ordered pcs in the chord progression. Here {\bf n} is an ordered vector of positive or negative integers where each component represents the minimal number of  steps that need to be applied to the corresponding pitches of the ordered pcs. So, for instance, VL(-1,-2,0) applied to [0,4,7] produces [7,11,2] (the I-V progression from the C major triad to its dominant in the key of C).\footnote{A full treatment of the mathematical properties of VL operators will be the subject of a forthcoming publication.}

\end{enumerate}

\section{\MUSICNTWRK}

The \MUSICNTWRK\ package (www.musicntwrk.com), is a python library written by the author and available as a PyPi project at www.pypi.org/project/music\-ntwrk/ or on GitHub: 
https://github.com/marcobn/musicntwrk.
\musicntwrk\ is the main module of the project and contains helper classes for pitch class set classification and manipulation in any arbitrary temperament ({\tt PCSet}, {\tt PCSetR} and {\tt PCSrow}), {\tt RHYTHMSeq} for the manipulation of rhythmic sequences, and the main class \musicntwrk\ that allows the construction of generalized musical space networks using distances between common descriptors (interval vectors, voice leadings, rhythm distance, etc.); the analysis of scores, the sonification of data and the generation of compositional frameworks. \musicntwrk\ acts as a wrapper for the various functions organized in the following sub-projects:
\begin{enumerate}
\item \networks\ - contains all the modules to construct dictionaries and networks of pitch class set spaces including voice leading, rhythmic spaces, timbral spaces and score network and orchestarion analysis 
\item \data\ - sonification of arbitrary data structures, including automatic score (musicxml) and MIDI generation
\item \timbre\ - analysis and characterization of timbre from a (psycho-)acoustical point of view. In particular, it provides: the characterization of sound using, among others, Mel Frequency or Power Spectrum Cepstrum Coefficients (MFCC or PSCC); the construction of timbral networks using descriptors based on MF- or PS-CCs
\item \harmony\ - helper functions for harmonic analysis, design and autonomous scoring
\item \mlutils\ - machine learning models for timbre recognition through the TensorFlow Keras framework
\item \plotting\ - plotting function including a module for automated network drawing
\item \utils\ - utility functions used by other modules
\end{enumerate}

\MUSICNTWRK\ is written in python 3 and requires installation of the following dependencies (done automatically when {\tt pip install musicntwrk}):
\footnote{this step might be unnecessary if running on a cloud service like Google Colaboratory.} 

\begin{enumerate}
\item System modules: {\tt sys, re, time, os}
\item Math modules: {\tt numpy, scipy}
\item Data modules: {\tt pandas, sklearn, networkx, python-louvain, tensorflow, powerlaw, ruptures, numba}
\item Music and audio modules: {\tt music21, librosa, pyo, pydub}
\item Visualization modules: {\tt matplotlib, vpython, PySimpleGUI}
\item Parallelization modules: {\tt mpi4py} (optional)
\end{enumerate}

The reader is encouraged to consult the documentation of each package to get acquainted with its purposes and use. In particular, \MUSICNTWRK\ relies heavily on the \texttt{music21} package for all the music theoretical and musicological functions.\cite{music21}
In what follows we provide the full API of \MUSICNTWRK\ only.
The display of musical examples in {\tt musicxml} format requires the installation of a score app like MuseScore (https://musescore.org/). See Section 08 of the {\tt music21} documentation for a step by step guide of installing a {\tt musicxml} reader.

Finally a full set of examples and application of the library for a variety of tasks can be downloaded from the \MUSICNTWRK\ repository on GitHub.
\footnote{https://github.com/marcobn/musicntwrk/tree/master/musicntwrk-2.0/examples}

\subsection{\musicntwrk}

\paragraph{The \texttt{PCSet} class.~}
The \texttt{PCSet} class deals with the classification and manipulation of pitch set classes generalized to arbitrary temperament systems (arbitrary number of pitches). The following methods are available:\\ \\
\texttt{def class PCSet}
\begin{itemize}
\item
  {\tt def \_\_init\_\_(self,pcs,TET=12,UNI=True,ORD=True)}
  \begin{itemize}
    {\item {{\tt pcs (int)}\ }}  pitch class set as list or numpy array
    {\item {{\tt TET (int)}\ }}  number of allowed pitches in the totality of the musical space (temperament). Default = 12 tones equal temperament
    {\item {{\tt UNI (logical)}\ }} if True, eliminate duplicate pitches (default)
    {\item {{\tt ORD (logical)}\ }} if True, sorts the pcs in ascending order (default)
  \end{itemize}

\item
    {\tt def normalOrder(self)}\\
        Order the pcs according to the most compact ascending scale in pitch-class space that spans less than an octave by cycling permutations.
\item 
    {\tt def normal0Order(self)}\\
        As normal order, transposed so that the first pitch is 0
\item
    {\tt def T(self,t=0)}\\
    Transposition by t (int) units (modulo TET)
\item
    {\tt def zeroOrder(self)}\\
        transposed so that the first pitch is 0
\item
	{\tt def M(self,t=1)}\\
		multiply pcs by an arbitrary scalar mod. 12
\item
	{\tt def multiplyBoulez(self,b)}\\
		pitch class multiplication of self * b according to P. Boulez
\item
    {\tt def I(self)}\\
        inverse operation: (-pcs modulo TET)
\item
    {\tt def primeForm(self)}\\
        most compact normal 0 order between pcs and its inverse
\item
    {\tt def intervalVector(self)}\\
        total interval content of the pcs
\item
    {\tt def LISVector(self)}\\
        Linear Interval Sequence Vector: sequence of intervals in an ordered pcs
\item
    {\tt def Op(self,name)}\\
    operate on the pcs with a distance operator
    \begin{itemize}
        {\item {{\tt name (str)}\ }} name of the operator O({ni})
    \end{itemize}
\item
    {\tt def VLOp(self,name)}\\
    operate on the pcs with a normal-ordered voice-leading operator 
    \begin{itemize}
        {\item {{\tt name (str)}\ }} name of the operator R({n$_0$,n$_1$,$...$,n$_{N_c}$})
    \end{itemize}
\item
    {\tt def forteClass(self)}\\
        Name of pcs according to the Forte classification scheme (only for TET=12)
\item
    {\tt def commonName(self)}\\
        Display common name of pcs (music21 function - only for TET=12)
\item
    {\tt def commonNamePrime(self)}\\
        As above, for prime forms
\item
    {\tt def commonNamePitched(self)}\\
        Name of chord with first pitch of pcs in normal order
\item
    {\tt def displayNotes(self,xml=False,prime=False)}\\
        Display pcs in score in musicxml format. If prime is True, display the prime form.
        \begin{itemize}
        {\item {{\tt xml (logical)}\ }} write notes on file in musicxml format
        {\item {{\tt prime (logical)}\ }} write pcs in prime form
    \end{itemize}
\end{itemize}

\paragraph{The \texttt{PCSetR} class.~}
The \texttt{PCSetR} class and its methods (listed below) parallels the \texttt{PCSet} class by adding recursive capabilities to it: in practice any method returns an instance of the class itself. This facilitates the construction of method chains for compositional or analytical tasks.  The following methods are available:\\ \\
\texttt{def class PCSetR}
\begin{itemize}
\item
  {\tt def \_\_init\_\_(self,pcs,TET=12,UNI=True,ORD=True)}
  \begin{itemize}
    {\item {{\tt pcs (int)}\ }}  pitch class set as list or numpy array
    {\item {{\tt TET (int)}\ }}  number of allowed pitches in the totality of the musical space (temperament). Default = 12 tones equal temperament
    {\item {{\tt UNI (logical)}\ }} if True, eliminate duplicate pitches (default)
    {\item {{\tt ORD (logical)}\ }} if True, sorts the pcs in ascending order (default)
  \end{itemize}

\item
    {\tt def normalOrder(self)}\\
        Order the pcs according to the most compact ascending scale in pitch-class space that spans less than an octave by cycling permutations.
\item 
    {\tt def normal0Order(self)}\\
        As normal order, transposed so that the first pitch is 0
\item
    {\tt def T(self,t=0)}\\
    Transposition by t (int) units (modulo TET)
\item
	{\tt def M(self,t=1)}\\
		multiply pcs by an arbitrary scalar mod. 12
\item
    {\tt def I(self)}\\
        inverse operation: (-pcs modulo TET)
\item
	{\tt def multiplyBoulez(self,b)}\\
		pitch class multiplication of self * b according to P. Boulez
\item
    {\tt def zeroOrder(self)}\\
        transposed so that the first pitch is 0
\item
    {\tt def inverse(self,pivot=0)}
    	invert pcs around a pivot pitch
\item
    {\tt def primeForm(self)}\\
        most compact normal 0 order between pcs and its inverse
\item
    {\tt def intervalVector(self)}\\
        total interval content of the pcs
\item
    {\tt def LISVector(self)}\\
        Linear Interval Sequence Vector: sequence of intervals in an ordered pcs
\item
    {\tt def Op(self,name)}\\
    operate on the pcs with a distance operator
    \begin{itemize}
        {\item {{\tt name (str)}\ }} name of the operator O({n$_i$})
    \end{itemize}
\item
    {\tt def VLOp(self,name)}\\
    operate on the pcs with a normal-ordered voice-leading operator 
    \begin{itemize}
        {\item {{\tt name (str)}\ }} name of the operator R({n$_0$,n$_1$,$...$,n$_{N_c}$})
    \end{itemize}
\item
    {\tt def NROp(self,ops=None)}\\
    operate on the pcs with a Neo-Rienmanian operator 
    \begin{itemize}
        {\item {{\tt ops (str)}\ }} name of the operator, P, L or R
    \end{itemize}
\item
	{\tt def opsNameVL(self,b,TET=12)}\\
		given a pcs returns the name of the normal-ordered voice-leading operator R that connects self to it
\item
	{\tt def opsNameO(self,b,TET=12)}\\
		given a pcs returns the name of the distance operator O that connects self to it
\end{itemize}

\paragraph{The {\tt PCSrow} class.~}
{\tt PCSrow} is a helper class for 12-tone rows operations (T,I,R,M,Q)\\ \\
\texttt{def class PCSrow}
\begin{itemize}
\item
  {\tt def \_\_init\_\_(self,pcs,TET=12)}
  \begin{itemize}
    {\item {{\tt pcs (int)}\ }}  12 tone row as  list or numpy array
  \end{itemize}
\item
    {\tt def normalOrder(self)}\\
        Transpose the row so that first pitch class is 0
\item
    {\tt def intervals(self)}\\
        vector of intervals of the row
\item 
	{\tt def T(self,t=0)}\\
	Transposition by t (int) units (modulo TET)
\item 
	{\tt def I(self)}\\
	inverse operation: (-pcs modulo TET)
\item 
	{\tt def R(self,t=1)}\\
	Retrograde plus transposition by t (int) units (modulo TET)
\item 
	{\tt def M(self,t=1)}\\
	Multiplication by t (int) units (modulo TET)
\item 
	{\tt def Q(self,t=1)}\\
	cyclic permutation of stride 6  so that the result is
	an All Interval Series in normal form
\item 
	{\tt star(self)}\\
	star of the row in prime form
\item 
	{\tt def constellation(self)}\\
	constellation of the row
\end{itemize}

\paragraph{The \texttt{RHYTHMSeq} class.~}
The {\tt RHYTHMSeq} class and its methods (listed below) encode various functions for rhythmic network manipulations. The {\tt RHYTHMSeq} class deals with the classification and manipulation of rhythmic sequences. The following methods are available:\\ \\ 
\texttt{def class RHYTHMSeq}
\begin{itemize}
\item
  {\tt \_\_init\_\_( self,rseq,REF='e',ORD=False)}
  \begin{itemize}
    {\item {{\tt rseq (str/fractions/floats)}\ }}  rhythm sequence as list of strings or fractions or floats
    {\item {{\tt REF (str)}\ }}  reference duration for prime form – the RHYTHMSeq class contains a dictionary of common duration notes that uses the fraction module for the definitions (implies  import fraction as fr):
    \begin{sloppypar}
{\tt\{'w': fr.Fraction(1,1),\ 'h': fr.Fraction(1,2),'q': fr.Fraction(1,4),\
 'e': fr.Fraction(1,8),\ 's': fr.Fraction(1/16),'t': fr.Fraction(1,32),\
 'wd': fr.Fraction(3,2),\ 'hd': fr.Fraction(3,4),'qd': fr.Fraction(3,8),\
 'ed': fr.Fraction(3,16),\ 'sd': fr.Fraction(3,32),'qt': fr.Fraction(1,6),\
 'et': fr.Fraction(1,12),\ 'st': fr.Fraction(1,24), 'qq': fr.Fraction(1,5),\
 'eq': fr.Fraction(1,10),\ 'sq': fr.Fraction(1,20)\}}.
 \end{sloppypar}
  This dictionary can be extended by the user on a case by case need.
    {\item {{\tt ORD (logical)}\ }} if {\tt True} sort durations in ascending order
      \end{itemize}
\item
    {\tt def normalOrder(self)}\\
        Order the rhythmic sequence according to the most compact ascending form.
\item
    {\tt def augment(self,t='e')}\\
        Augmentation by t units
        \begin{itemize}
        {\item {{\tt t (str)}\ }} duration of augmentation
        \end{itemize}
\item
    {\tt def diminish(self,t='e')}\\
        Diminution by t units
        \begin{itemize}
        {\item {{\tt t (str)}\ }} duration of diminution
        \end{itemize}
\item
    {\tt def retrograde(self)}\\
        Retrograde operation
\item
    {\tt def isNonRetro(self)}\\
        Check if the sequence is not retrogradable
\item
    {\tt def primeForm(self)}\\
        reduce the series of fractions to prime form        
\item
    {\tt def durationVector(self,lseq=None)}\\
        total relative duration ratios content of the sequence
        \begin{itemize}
        {\item {{\tt lseq (list of fractions) }\ }} reference list of duration for evaluating interval content; the default list is:
		\begin{sloppypar}
		{\tt fr.Fraction(1/8), fr.Fraction(2/8), fr.Fraction(3/8),
 			fr.Fraction(4/8), fr.Fraction(5/8), fr.Fraction(6/8),
 			fr.Fraction(7/8), fr.Fraction(8/8), fr.Fraction(9/8)}
		\end{sloppypar}
    \end{itemize}
\item
    {\tt def durationVector(self,lseq=None)}\\
        inter-onset duration interval content of the sequence
        \begin{itemize}
        {\item {{\tt lseq (list of fractions) }\ }} reference list of duration for evaluating interval content; the default list is the same as above.
        \end{itemize}

\end{itemize}

\paragraph{The \musicntwrk\ class}
Defines wrappers around calls to the main functions of the different packages. The variables passed in the {\tt def}s are used to call the 
specific function requested. See documentation of the functions in the separate sections.
\begin{itemize}
\item
	\begin{sloppypar}
    {\tt def dictionary(self, space=None, N=None, Nc=None, order=None, row=None, a=None, prob=None, REF=None, scorefil=None, music21=None, show=None)}
    \end{sloppypar}
	define dictionary in the musical space specified in 'space'.
	All other variables are as defined in {\tt pcsDictionary}, {\tt rhythmDictionary}, {\tt orchestralVector}, and {\tt scoreDictionary}, 
	in Sec. \ref{networks} depending on the choice of 'space'
	\begin{itemize}
		\begin{sloppypar}
        {\item {{\tt space (string)}\ }} = 'pcs', pitch class sets dictionary; 'rhythm' or 'rhythmP', rhythm dictionaries; 'score', score dictionary;
         and 'orch' orchestral vector.
        \end{sloppypar}
    \end{itemize}
	{\it Returns}\\
	See description in Sec. \ref{networks} for individual functions.
\item
	\begin{sloppypar}
    {\tt def network(self, space=None, label=None, dictionary=None, thup=None, thdw=None, thup\_e=None, thdw\_e=None, distance=None, prob=None, write=None, pcslabel=None, vector=None, ops=None, name=None, ntx=None, general=None, seq=None, sub=None, start=None, end=None, grphtype=None, wavefil=None, cepstrum=None, color=None)}
	\end{sloppypar}
	define networks in the musical space specified in 'space': 
	\begin{itemize}
        {\item {{\tt space (string)}\ }} = 'pcs', pitch class sets network, both full and ego network from a given pcs; 
        'rhythm' or 'rhythmP', rhythm dictionaries (see below for details); 'score', score dictionary;
         and 'orch' orchestral vector. See description in Sec. \ref{networks}.
    \end{itemize}
	{\it Returns}\\
	See description in Sec. \ref{networks} for individual functions.

\item
    {\tt def timbre(self, descriptor=None, path=None, wavefil=None, standard= None,
   nmel=None, ncc=None, zero=None, lmax=None, maxi= None, nbins = None, method=None,
    scnd=None, nstep=None)}\\
	Define sound descriptors for timbral analysis: MFCC, PSCC ASCBW in regular or standardized form. See description of variables in Sec. \ref{timbre}.
\item
    {\tt def harmony(self,descriptor=None,mode=None,x=None,y=None)}\\
	handler for calculating tonal harmony models, tonnentz and to launch the tonal harmony calculator. See description of variables in Sec. \ref{harmony}.
\item
\begin{sloppypar}
    {\tt def sonify(self, descriptor=None, data=None, length=None, midi=None, scalemap=None,
    ini=None, fin=None, fac=None, dur=None, transp=None, col=None, write=None, vnorm=None,
    plot=None, crm=None, tms=None, xml=None)}
    \end{sloppypar}
	sonification strategies - simple sound (spectral) or score (melodic progression). See description of variables in Sec. \ref{data}
\end{itemize}

\subsection{\networks}
\label{networks}
\networks\ contains specific functions for network generation and analysis in different musical spaces.\\

Networks of pitch class sets

\begin{itemize}
\item
    {\tt def pcsDictionary(Nc,order=0,TET=12,row=False,a=None)}\\
        Generate the dictionary of all possible pcs of a given cardinality in a generalized musical space of TET pitches. Returns the dictionary as pandas DataFrame and the list of all Z-related pcs
        \begin{itemize}
        {\item {{\tt Nc (int)}\ }} cardinality
        {\item {{\tt order (logical)}\ }} if 0 returns pcs in prime form, if 1 retrns pcs in normal order, if 2, returns pcs in normal 0 order
        {\item {{\tt row (logical)}\ }} if True build dictionary from tone row, if False, build dictionary from all combinatorial pcs of Nc cardinality 
        given the totality of TET.
        {\item {{\tt a (int)}\ }} if row = True, a is the list of pitches in the tone row
    \end{itemize}
	{\it Returns}
	\begin{itemize}
        {\item {{\tt dictionary (pandas dataframe object)}\ }} dictionary as dataframe (name, pitch class set, interval vector)
        {\item {{\tt ZrelT (list of strings)}\ }} pitch class sets that have a Z-relation
    \end{itemize}

\item
    {\tt def pcsNetwork(input\_csv, thup= 1.5, thdw=0.0,TET=12, distance=\\ 'euclidean', col=2,prob=1)}\\
        generate the network of pcs based on distances between interval vectors\\
        In output it writes the nodes.csv and edges.csv as separate files in csv format
        \begin{itemize}
        {\item {{\tt input\_csv (str)}\ }} file containing the dictionary generated by pcsNetwork
        {\item {{\tt thup, thdw (float)}\ }} upper and lower thresholds for edge creation
        {\item {{\tt distance (str)}\ }} choice of norm in the musical space, default is 'euclidean'
        {\item {{\tt col (int)}\ }} metric based on interval vector, col = 1 can be used for voice leading networks in spaces of fixed cardinality (OBSOLETE)
        {\item {{\tt prob (float)}\ }} if not 1, defines the probability of acceptance of any given edge
    \end{itemize}
	{\it Returns}
	\begin{itemize}
        {\item {{\tt nodes, edges (pandas dataframe objects)}\ }} dataframe of nodes and edges of the network
    \end{itemize}    

\item
    {\tt def pcsEgoNetwork(label, input\_csv, thup\_e=5.0, thdw\_e=0.1, thup= 1.5, thdw=0.1, TET=12, distance='euclidean')}\\
        Generates the network for a focal node (ego) and the nodes to whom ego is directly connected to (alters). In output it writes the nodes\_ego.csv, edges\_ego.csv and edges\_alters.csv as separate files in csv format
        \begin{itemize}
        {\item {{\tt label (str)}\ }} label of the ego node
        {\item {{\tt thup\_e, thdw\_e (float)}\ }} upper and lower thresholds for edge creation from ego node
        {\item {{\tt thup, thdw (float)}\ }} upper and lower thresholds for edge creation among alters
        {\item {{\tt distance (str)}\ }} choice of norm in the musical space, default is 'euclidean'
    \end{itemize}
	{\it Returns}
	\begin{itemize}
        {\item {{\tt nodes(ego), edges(ego), edges(alters) (pandas dataframe objects)}\ }} data\-frames of nodes and edges of the  ego network
    \end{itemize}      
\item
    {\tt def vLeadNetwork(input\_csv,thup=1.5,thdw=0.1,TET=12,w=True,\\ distance='euclidean',prob=1)}\\
        Generation of the network of all minimal voice leadings in a generalized musical space of TET pitches - based on the minimal distance operators - select by distance. In output returns nodes and edges tables as pandas DataFrames.
        \begin{itemize}
        {\item {{\tt input\_csv (str)}\ }} file containing the dictionary generated by pcsNetwork
        {\item {{\tt thup, thdw (float)}\ }} upper and lower thresholds for edge creation
        {\item {{\tt distance (str)}\ }} choice of norm in the musical space, default is 'euclidean'
        {\item {{\tt w (logical)}\ }} if True it writes the nodes.csv and edges.csv files in csv format
    \end{itemize}
	{\it Returns}
	\begin{itemize}
        {\item {{\tt nodes, edges (pandas dataframe objects)}\ }} dataframe of nodes and edges of the network
    \end{itemize}      
\item
    {\tt def vLeadNetworkByName(input\_csv, thup=1.5, thdw=0.1, TET=12, w=True, distance= 'euclidean', prob=1)}\\
        Generation of the network of all minimal voice leadings in a generalized musical space of TET pitches - based on the minimal distance operators - select by name. In output returns nodes and edges tables as pandas DataFrames. Available also in vector form for computational efficiency as {\tt vLeadNetworkByNameVec}
        \begin{itemize}
        {\item {{\tt input\_csv (str)}\ }} file containing the dictionary generated by pcsNetwork
        {\item {{\tt name (str)}\ }} name of operator for edge creation
        {\item {{\tt distance (str)}\ }} choice of norm in the musical space, default is 'euclidean'
        {\item {{\tt w (logical)}\ }} if True it writes the nodes.csv and edges.csv files in csv format
        \end{itemize}
	{\it Returns}
	\begin{itemize}
        {\item {{\tt nodes, edges (pandas dataframe objects)}\ }} dataframe of nodes and edges of the network
    \end{itemize}  
\end{itemize}
    
Score networks        
 
\begin{figure}
\includegraphics[width=\textwidth]{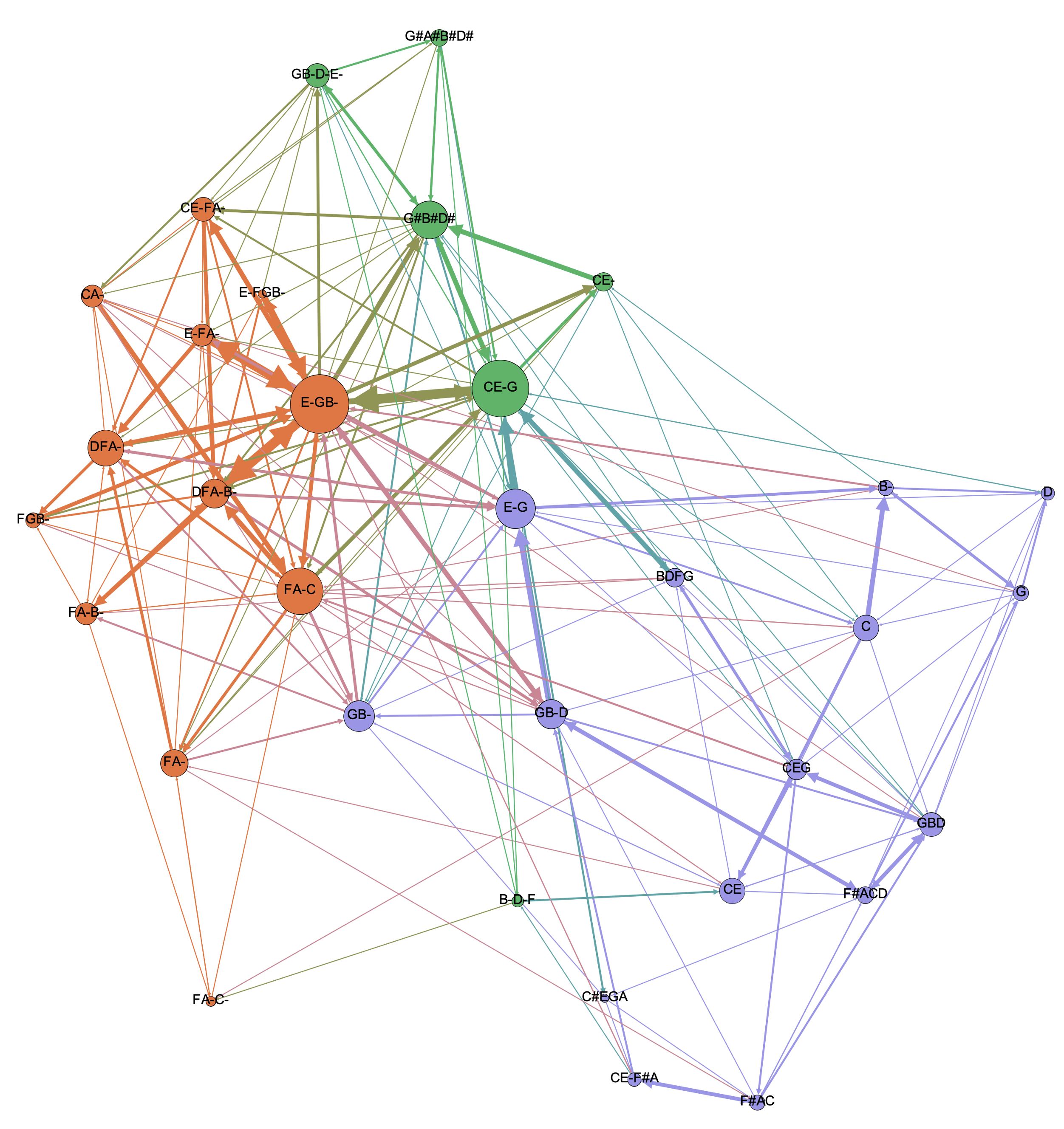}
\caption{Score network of the first movement of L. van Beethoven string quartet Op. 127 n. 12 in E$\flat$ Major. Different coloring of the nodes displays their modularity class and reflects the tonal regions used by Beethoven in this piece.}\label{LvBscore}
\end{figure}

\begin{itemize}
\item
    {\tt def scoreNetwork(seq, ntx, general, distance, TET)}\\
        Generates the directional network of chord progressions from any score in musicxml format, See Figure \ref{LvBscore} for an illustration.
        \begin{itemize}
        {\item {{\tt seq (int)}\ }} list of pcs for each chords extracted from the score
        {\item {{\tt ntx (logical)}\ }} if {\tt True} produces the network (directed and undirected) as a {\tt networkx} graph object
        {\item {{\tt general (logical)}\ }} if {\tt True} classifies the chord progressions with the voice-leading operators, else with the distance operators
        {\item {{\tt distance (string)}\ }} defines the metric using the definitions in {\tt sklearn}.
        \end{itemize}
    {\it Returns}
	\begin{itemize}
        {\item {{\tt nodes, edges (pandas dataframe objects)}\ }} dataframe of nodes and edges of the network
        {\item {{\tt counts (list of integers)}\ }} number of occurrences for each chord (histogram)
        {\item {{\tt avgdeg, modul (floats)}\ }} if {\tt ntx = True}: average degree and modularity
        {\item {{\tt Gx, Gxu (networkx graph objects)}\ }} if {\tt ntx = True}: directed and undirected graph as {\tt networkx} objects
    \end{itemize} 
\item
    {\tt def scoreSubNetwork(seq, start, end, ntx, general, distance, TET)}\\
        Generates the directional sub- network of chord progressions from a range of pcs in the sequence defined by {\tt start} and {\tt stop} (int). All
        other variables as above. It returns the sub-network in the same format as above..
\item
    {\tt def scoreDictionary(seq, TET=12)}\\
        Builds the dictionary of pcs in any score in musicxml format
	\begin{itemize}
        {\item {{\tt seq (list of lists)}\ }} list of the pitch class set sequence (from {\tt readScore})
    \end{itemize}
    {\it Returns}
	\begin{itemize}
        {\item {{\tt dictionary (pandas dataframe object)}\ }} dictionary as dataframe (name, pitch class set, interval vector)
    \end{itemize}
\item
    {\tt def readScore(inputxml,TET=12,music21=False)}\\
        Reads musicxml score and returns chord sequence
        \begin{itemize}
        {\item {{\tt inputxml (str)}\ }} score file
        {\item {{\tt music21 (logical)}\ }} if True search the music21 corpus
        \end{itemize}
	{\it Returns}
	\begin{itemize}
        {\item {{\tt seq (list of lists)}\ }} list of the pitch class set sequence
        {\item {{\tt chords (music21 chord object)}\ }} list of the chord sequence as extracted by {\tt music21}
    \end{itemize}
            
\end{itemize}

Rhythm networks

\begin{itemize}
\item
    {\tt def rhythmDictionary(Nc,a=None,REF='e')}\\
        Generates the dictionary of all possible rhythmic sequences of Nc length in a generalized meter space of N durations. Returns the dictionary as pandas DataFrame and indicates all non retrogradable and Z-related cells
        \begin{itemize}
        {\item {{\tt Nc (int)}\ }} cell length
        {\item {{\tt a (str)}\ }} list of durations in the rhythm sequence
    \end{itemize}
	{\it Returns}
	\begin{itemize}
        {\item {{\tt dictionary (pandas dataframe object)}\ }} dictionary as dataframe (name, rhythm sequence, duration vector)
        {\item {{\tt ZrelT (list of strings)}\ }} rhythm sequences that have a Z-relation
    \end{itemize}
\item
    {\tt def rhythmPDictionary(N,Nc,REF='e')}\\
        Generate the dictionary of all possible rhythmic sequences from all possible groupings of N REF durations. Returns the dictionary as pandas DataFrame and indicates all non retrogradable and Z-related cells
        \begin{itemize}
        {\item {{\tt Nc (int)}\ }} cell length
        {\item {{\tt N (int)}\ }} number of REF units
    \end{itemize}
	{\it Returns}\\
	As above.
\item
    {\tt def rhythmNetwork(input\_csv, thup=1.5, thdw=0.0, distance= 'eucl\-idean', prob=1, w=False)}\\
        Generates the network of rhythmic cells based on distances between duration vectors.
        In output it writes the nodes.csv and edges.csv as separate files in csv format
        \begin{itemize}
        {\item {{\tt input\_csv (str)}\ }} file containing the dictionary generated by rhythmNetwork
        {\item {{\tt thup, thdw (float)}\ }} upper and lower thresholds for edge creation
        {\item {{\tt distance (str)}\ }} choice of norm in the musical space, default is 'euclidean'
        {\item {{\tt prob (float)}\ }} if not 1, defines the probability of acceptance of any given edge
         {\item {{\tt w (logical)}\ }} if True it writes the nodes.csv and edges.csv files in csv format
    \end{itemize}
	{\it Returns} 
	\begin{itemize}
        {\item {{\tt nodes, edges (pandas dataframe objects)}\ }} dataframe of nodes and edges of the network
    \end{itemize} 
\item
    {\tt def rLeadNetwork(input\_csv, thup=1.5, thdw=0.1, w=True, distance= 'euclidean', prob=1)}\\
        Generation of the network of all minimal rhythm leadings in a generalized musical space of N$_C$-dim rhythmic cells based on the rhythm distance operator. Returns nodes and edges tables as pandas DataFrames
        \begin{itemize}
        {\item {{\tt input\_csv (str)}\ }} file containing the dictionary generated by rhythmNetwork
        {\item {{\tt thup, thdw (float)}\ }} upper and lower thresholds for edge creation
        {\item {{\tt distance (str)}\ }} choice of norm in the musical space, default is 'euclidean'
        {\item {{\tt prob (float)}\ }} if not 1, defines the probability of acceptance of any given edge
         {\item {{\tt w (logical)}\ }} if True it writes the nodes.csv and edges.csv files in csv format
    \end{itemize}
	{\it Returns}\\
	As above.
\end{itemize}

Orchestration networks

\begin{figure}
\includegraphics[width=\textwidth]{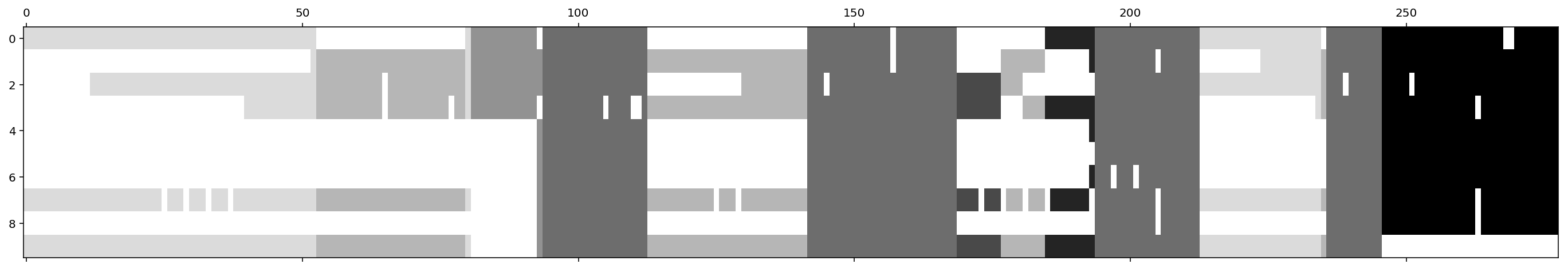}
\caption{Orchestration map of the first movement (Allegro) of J.S. Bach's Brandenburg Concerto n. 2, BWV 1047, as produced by the {\tt orchestralVectorColor} function. Different shades of gray represent the sections of similar orchestral color as measured by their modularity class in the network.} \label{BachOrch}
\end{figure}

\begin{itemize}
\item
    {\tt def orchestralVector(inputfile,barplot=True)}\\
        Builds the orchestral vector sequence from score in {\tt musicxml} format. Returns the score sliced by beat; orchestration vector.
        \begin{itemize}
        {\item {{\tt barplot=True}\ }} plot the orchestral vector sequence as a matrix
    \end{itemize}
	{\it Returns} 
	\begin{itemize}
        {\item {{\tt score (music21 object)}\ }} score sliced by beat
        {\item {{\tt orch (list of lists)}\ }} orchestral vectors for every beat (in binary format: 1 if instrument is playin, 0 if not)
        {\item {{\tt num (list)}\ }} beat vector identifier (integer corresponding to the binary number encoded in the orchestral vector)
    \end{itemize} 
\item
    {\tt def orchestralNetwork(seq)}\\
        Generates the directional network of orchestration vectors from any score in musicxml format.
        Use {\tt orchestralVector} to import the score data as sequence. Returns nodes and edges as Pandas DataFrames; average degree, modularity and partitioning of the network. 
        \begin{itemize}
        {\item {{\tt seq (int)}\ }} list of orchestration vectors extracted from the score
    \end{itemize}
	{\it Returns}\\
	As {\tt scoreNetwork}.
\item
    {\tt def orchestralVectorColor(orch, dnodes, part, color=plt.cm.binary)}\\
        Plots the sequence of the orchestration vectors color-coded according to the modularity class they belong. Requires the output of orchestralNetwork(). 
        See Figure \ref{BachOrch}
        \begin{itemize}
        {\item {{\tt orch (int)}\ }} list of orchestration vectors extracted from the score
        {\item {{\tt dnodes (pandas dataframe object)}\ }} dataframe of nodes of the network (for labelling)
        {\item {{\tt part (list of int)}\ }} list of partitions for orchestration vectors
    \end{itemize}    

\end{itemize}

Timbral networks

\begin{itemize}
\item
    {\tt def timbralNetwork(waves,vector,thup=10,thdw=0.1)}\\
        generates the network of MFCC vectors from sound recordings. Returns the nodes and edges tables as pandas DataFrames
        \begin{itemize}
        {\item {{\tt seq (float)}\ }} list of MFCC vectors 
        {\item {{\tt waves (str)}\ }} names of sound files
    \end{itemize}
	{\it Returns} 
	\begin{itemize}
        {\item {{\tt nodes, edges (pandas dataframe objects)}\ }} dataframe of nodes and edges of the network
    \end{itemize}     
\end{itemize}
See Sec. \ref{timbre} for a full discussion on the {\tt timbre} modules.

\subsection{\timbre}\label{timbre}
\timbre\ contains all the modules that deal with analysis and characterization of timbre from a (psycho-)acoustical point of view and provides the characterization of sound using, among others, Mel Frequency or Power Spectrum Cepstrum Coefficients (MFCC or PSCC) that can be used in the construction of timbral networks using these descriptors

\paragraph{Sound classification.~}
The sound classification section, {\tt timbre}, comprises of modules for specific sound analysis that are based on the {\tt librosa} python library for audio signal processing. We refer the interested reader to the {\tt librosa} documentation at {\tt https://librosa.github.io/librosa/index.html}. For a more complete discussion on the descriptors defined in \MUSICNTWRK\ please refer to the work in Ref. ~\cite{ref_article4}.

\paragraph{Audio descriptors and metrics in the generalized timbre space.}
The Power Cepstrum of a signal gives the rate of change of the envelope of different spectrum bands and is defined as the squared magnitude of the inverse Fourier transform of the logarithm of the squared magnitude of the Fourier transform of a signal:
$$
\textrm{PSCC} = \left | FT^{-1} \left\{ \log (|FT\{f(t)\right\}|^2)\right |^2
$$
\MUSICNTWRK\ defaults to the first 13 cepstrum coefficients (PSCC), where the 0$^{th}$ coefficient corresponds to the power distribution of the sound over time. The number of coefficient can be controlled by the user.

The Mel Frequency Cepstrum of a signal is obtained as in the equation for the PSCC and differ from the power cepstrum by the choice of the spectrum bands that are mapped over the Mel scale using triangular overlapping windows.  The mapping of the frequency bands on the Mel scale better approximates the human auditory system's response than the linearly-spaced frequency bands used in the normal cepstrum. \MUSICNTWRK\ defaults to  16 bands Mel filter. As for the PSCC, the 0-th coefficient corresponds to the power distribution of the sound over time. Both PSCC and MFCC are obtained using 64 bins in the short time Fourier transform (default).

\begin{figure}
\centering
\vspace{0pt}
\includegraphics[scale=0.40]{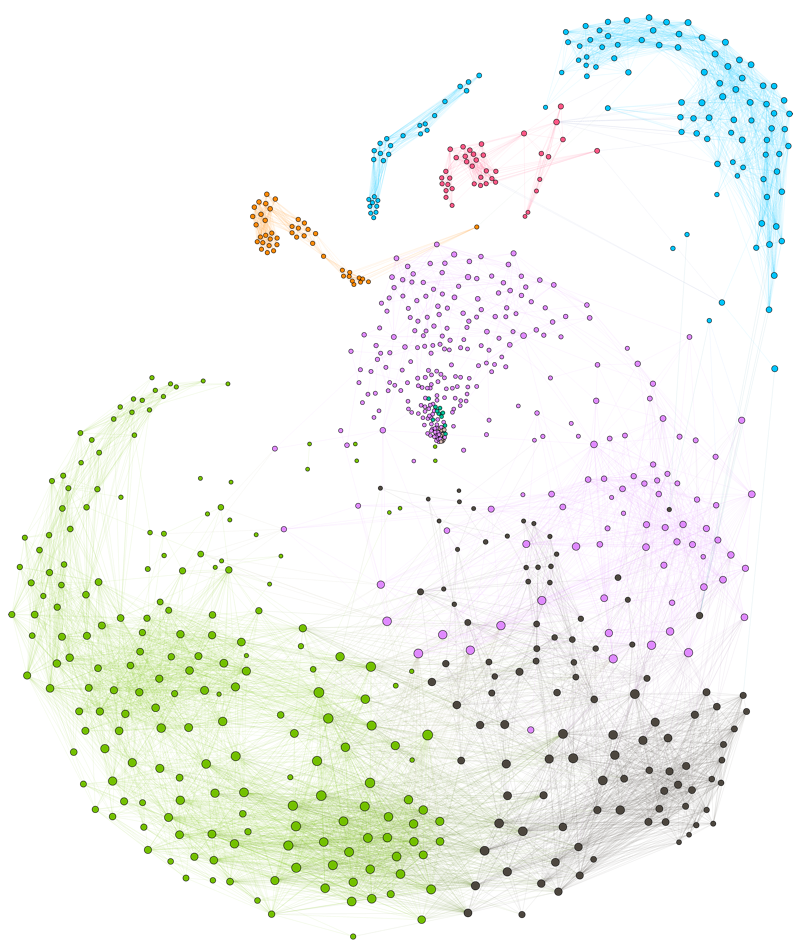}
\caption{Section of the network of the MFCCs built from the 1620 sounds that have been used to train the machine learning models. Colors indicate the classification based on their modularity class: \textcolor{Mygreen}{Green}, mostly high frequency tones from wood;
\textcolor{Mygrey}{Gray}, mostly high to mid-frequency tones of wood; \textcolor{Mypurple}{Purple}, mostly deep frequency tones of wood; \textcolor{Mycyan}{Cyan}, mostly  dry metal tones; \textcolor{Mypink}{Pink}, mostly metal; and \textcolor{Myorange}{Orange}, mostly  "choked" metal sounds.}
\label{modelnet}
\end{figure}

As an illustration, in Figure \ref{modelnet} we show the network of the MFCC built from 1620 sounds produced by either wooden or metallic objects that were analyzed by a deep learning model trained on a perceptually salient acoustic descriptor or on a signal descriptor based on the energy contents of the signal for a study on automatic timbre characterization.\cite{ref_article4} In the figure we display the principal component of the timbral network for which less than 2\% of all possible edges are built, that is, we allow an edge only if two MFCCs are at a distance that is less than 3\% of the maximum diameter of the network. This representation reveals that the classification is coherent with material categories, as it can be observed from the emergence of clusters of sounds belonging to the same material (see for instance the green cluster of wood sounds on the lower left part of the network and the cyan, pink and orange clusters of metallic sounds in the upper part). This result validates the corpus with respect to the sound quality. 

\begin{itemize}
\item
    {\tt def computeMFCC(input\_path,input\_file,barplot=True,zero=True)}\\
        read audio files in repository and compute a normalized MEL Frequency Cepstrum Coefficients and single vector map of the full temporal evolution of the sound as the convolution of the time-resolved MFCCs convoluted with the normalized first MFCC component (power distribution). Returns the list of files in repository, MFCC0, MFCC coefficients.
        \begin{itemize}
        {\item {{\tt input\_path (str)}\ }} path to repository
        {\item {{\tt input\_file (str)}\ }} filenames (accepts "*")
        {\item {{\tt barplot (logical)}\ }} plot the MFCC0 vectors for every sound in the repository 
        {\item {{\tt zero (logical)}\ }} If False, disregard the power distribution component. 
    \end{itemize}
	{\it Returns} 
	\begin{itemize}
        {\item {{\tt waves (list of strings)}\ }} filenames of the .wav files in the repository
        {\item {{\tt mfcc0 (list)}\ }} vector of MFCC0 (MFCC with DC component taken out)
        {\item {{\tt mfcc (list)}\ }} vector of MFCC (full MFCC)
    \end{itemize}    
\item
    {\tt def computePSCC(input\_path,input\_file,barplot=True,zero=True)}\\
        Reads audio files in repository and compute a normalized Power Spectrum Frequency Cepstrum Coefficients and single vector map of the full temporal evolution of the sound as the convolution of the time-resolved PSCCs convoluted with the normalized first PSCC component (power distribution). Returns the list of files in repository, PSCC0, PSCC coefficients. Other variables and output as above.
\item
    {\tt def computeStandardizedMFCC(input\_path,input\_file,nmel=16,\\ nmfcc=13,lmax=None,nbins=None)}\\
        read audio files in repository and compute the standardized (equal number of samples per file) and normalized MEL Frequency Cepstrum Coefficient. Returns the list of files in repository, MFCC coefficients, standardized sample length.
        \begin{itemize}
        {\item {{\tt nmel (int)}\ }} number of Mel bands to use in filtering 
        {\item {{\tt nmfcc (int)}\ }} number of MFCCs to return
        {\item {{\tt lmax (int)}\ }} max number of samples per file
        {\item {{\tt nbins (int)}\ }} number of FFT bins
    \end{itemize}
    {\it Returns}\\
    As above.
\item
    {\tt def computeStandardizedPSCC(input\_path,input\_file,nmel=16,\\ psfcc=13,lmax=None,nbins=None)}\\
        read audio files in repository and compute the standardized (equal number of samples per file) and normalized Power Spectrum Frequency Cepstrum Coefficients. Returns the list of files in repository, PSCC coefficients, standardized sample length.\\
        Variables defined as for MFCCs.
    {\it Returns}\\
    As above.

\end{itemize}

\begin{figure}
\includegraphics[width=\textwidth]{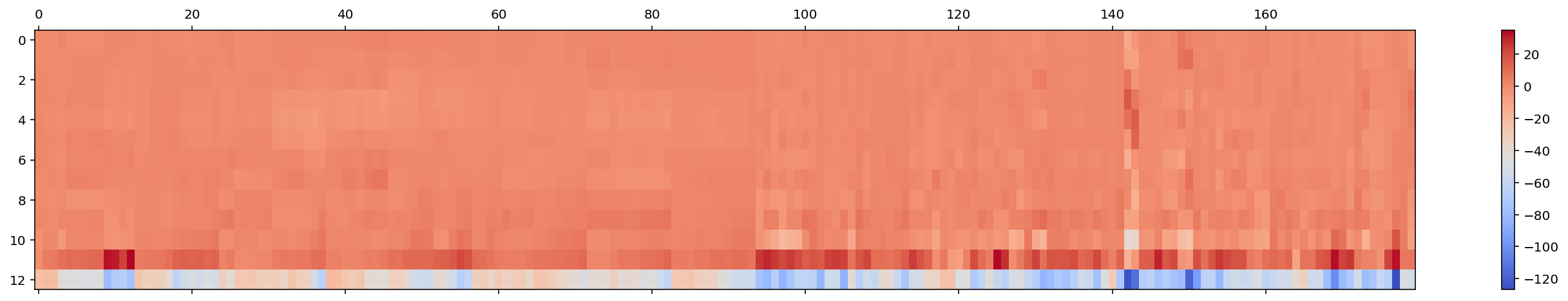}
\caption{Map of the MFCC for a repository of 180 impact sounds.} \label{PSCC}
\end{figure}

\subsection{Machine Learning Models ({\tt ml\_utils})} 
\begin{figure}
\includegraphics[width=\textwidth]{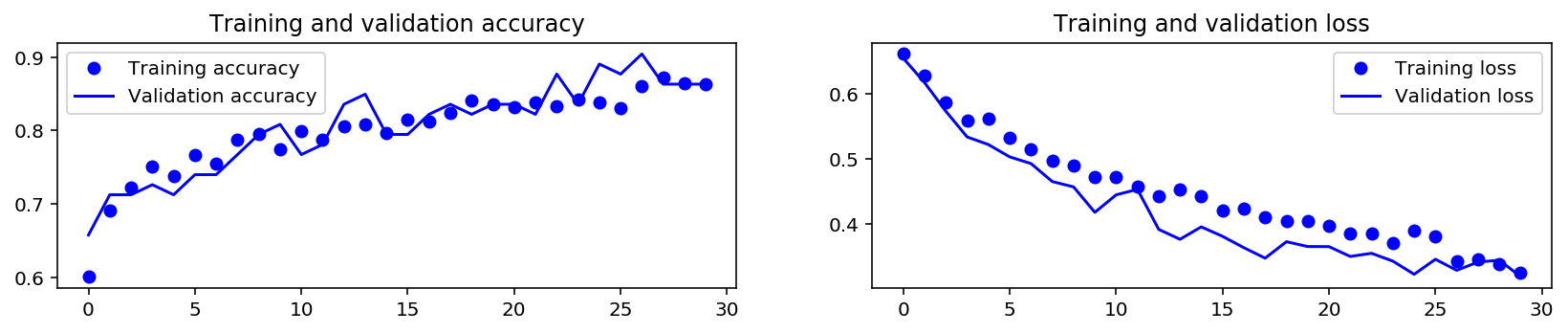}
\caption{Training and validation accuracy and loss in a typical Neural Network model learning run} \label{accloss}
\end{figure}

The definition of machine learning models for sound recognition requires standard techniques of data science (like the separation of data entries in training and testing sets, definition of neural network architectures, etc.) that will not be discussed here. Basic knowledge of Keras is also assumed. {\tt ml\_utils} contains many auxiliary functions to deal with such tasks. We implemented a deep learning model based on convolutional neural network (CNN) architecture inspired by similar approaches used in image and sound recognition \cite{piczak2015}. The CNN is built using the Keras kernel of Tensorflow \cite{tensorfl} and it is trained on the full PSCC or MFCC data, after proper scaling and normalization. We retained only models with validation accuracy higher than 90\%. After an appropriate model is chosen, it is tested on the set initially chosen for the human perception experiment. Each model chosen retains a similar accuracy on this set. A typical result of a training session on 30 epochs is shown in Fig. \ref{accloss}.
Here we limit to report the API for the main machine learning functions and refer the user to the full example available of GitHub:

\begin{itemize}
\item
    {\tt def trainNNmodel(mfcc, label, gpu=0, cpu=4, niter=100, nstep = 10,  
    neur = 16, test=0.08, num\_classes=2, epoch=30, verb=0, thr=0.85, w=False)}\\
        train a 2 layer neural network model on the ful MFCC spectrum of sounds. Returns: model,training and testing sets,data for re-scaling and normalization,data to asses the accuracy of the training session.
        \begin{itemize}
        {\item {{\tt mfcc (float)}\ }} list of all the MFCCs (or PSCCs) in the repository 
        {\item {{\tt gpu, cpu (int)}\ }} number of GPUs or CPUs used for the run
        {\item {{\tt niter (int)}\ }} max number of model fit sessions
        {\item {{\tt nstep (int)}\ }} how often the training and testing sets are redefined
        {\item {{\tt neur (int)}\ }} number of neurons in first layer (it is doubled on the second layer
        {\item {{\tt test (float)}\ }} defines the relative size of training and testing sets
        {\item {{\tt num\_classes=2 (int)}\ }} dimension of the last layer
        {\item {{\tt epoch (int)}\ }} number of epochs in the training of the neural network
         {\item {{\tt verb (int)}\ }} verbose - print information during the training run
        {\item {{\tt thr (float)}\ }} keep the model if accuracy is > test
        {\item {{\tt w (logical)}\ }} write model on file if accuracy is above {\tt thr}
    \end{itemize}
\item
    {\tt def trainCNNmodel(mfcc, label, gpu=0, cpu=4, niter=100, nstep=10,   
    neur=16, test=0.08, num\_classes=2, epoch=30, verb=0, thr=0.85, w = False)}\\
        train a convolutional neural network (CNN) model on the full MFCC/PSCC spectrum of sounds. Returns: model,training and testing sets,data for re-scaling and normalization,data to asses the accuracy of the training session.\\
        Parameters are defined as above.
\end{itemize}
For a complete description and example see the notebook on GitHub.

\subsection{\data}\label{data}
\data\ contains functions for the sonification of data in multi-column or csv format and produces output as WAV either via csound (it requires an installation of csound and direct reference to the {\tt ctcsound} module), {\tt pyo} or pure python, and {\tt musicxml} or MIDI. Two sonification protocols are available: spectral - data are mapped to a single sound using subtractive synthesis (FIR filter); and linear - individual data points are mapped to pitches in a time-series structure. See Ref. ~\cite{ref_article5,ref_article3b} for a complete description of this protocol. \data\ contains:

\begin{itemize}
\item
    {\tt def r\_1Ddata(path,fileread)}\\
        Read data file in a multicolumn format (csv files can be easily put in this format using Pandas). Returns the data values as (x,y).
        \begin{itemize}
        {\item {{\tt path (str)}\ }} path to data file 
        {\item {{\tt fileread (str)}\ }} data file
    \end{itemize}
\item
    {\tt def i\_spectral2(xv,yv,itime,path='./',instr='noise')}\\
        Use subtractive synthesis to sonify data structure. Returns the sound file.
        \begin{itemize}
        {\item {{\tt xv,yv (float)}\ }} data structure to sonify
        {\item {{\tt path (str)}\ }} path to data file 
        {\item {{\tt fileread (str)}\ }} data file
    \end{itemize}
\item
    {\tt def i\_time\_series(xv,yv,path='./',instr='csb701')}\\
        Use csound instruments to sonify data structures as time-series.  Returns the sound file.
        \begin{itemize}
        {\item {{\tt xv,yv (float)}\ }} data structure to sonify
        {\item {{\tt path (str)}\ }} path to data file 
        {\item {{\tt fileread (str)}\ }} data file
        {\item {{\tt instr (str)}\ }} csound instrument (it can be modified by user)
    \end{itemize}
\item
    {\tt def MIDImap(pdt,scale,nnote)}\\
        Data to MIDI conversion on a given scale defined in scaleMapping (see below). Returns the MIDI data structure.
        \begin{itemize}
        {\item {{\tt pdt (float)}\ }} data structure mapped to MIDI numbers
        {\item {{\tt scale (float)}\ }} scale mapping (from scaleMapping) 
        {\item {{\tt nnote (int)}\ }} number of notes in the scale (from scaleMapping)
    \end{itemize}
\item
    {\tt def scaleMapping(scale)}\\
        Scale definitions for MIDI mapping. Returns: scale, nnote (see above).
\item
    {\tt def MIDIscore(yvf,dur=2,w=None,outxml='./music',outmidi='./music')}\\
        Display score or writes to file
        \begin{itemize}
        {\item {{\tt yvf (float)}\ }} data structure mapped to MIDI numbers (from MIDImap)
        {\item {{\tt dur (int)}\ }} reference duration 
        {\item {{\tt w (logical)}\ }} if True writes either musicxml or MIDI file)
    \end{itemize}
\item
    {\tt def MIDImidi(yvf,vnorm=80,dur=4,outmidi='./music')}\\
        Display score or writes to file
        \begin{itemize}
        {\item {{\tt yvf (float)}\ }} data structure mapped to MIDI numbers (from MIDImap)
        {\item {{\tt vnorm (int)}\ }} reference velocity 
        {\item {{\tt outmidi (str)}\ }} MIDI file
    \end{itemize}
\end{itemize}

\begin{figure}
\includegraphics[width=\textwidth]{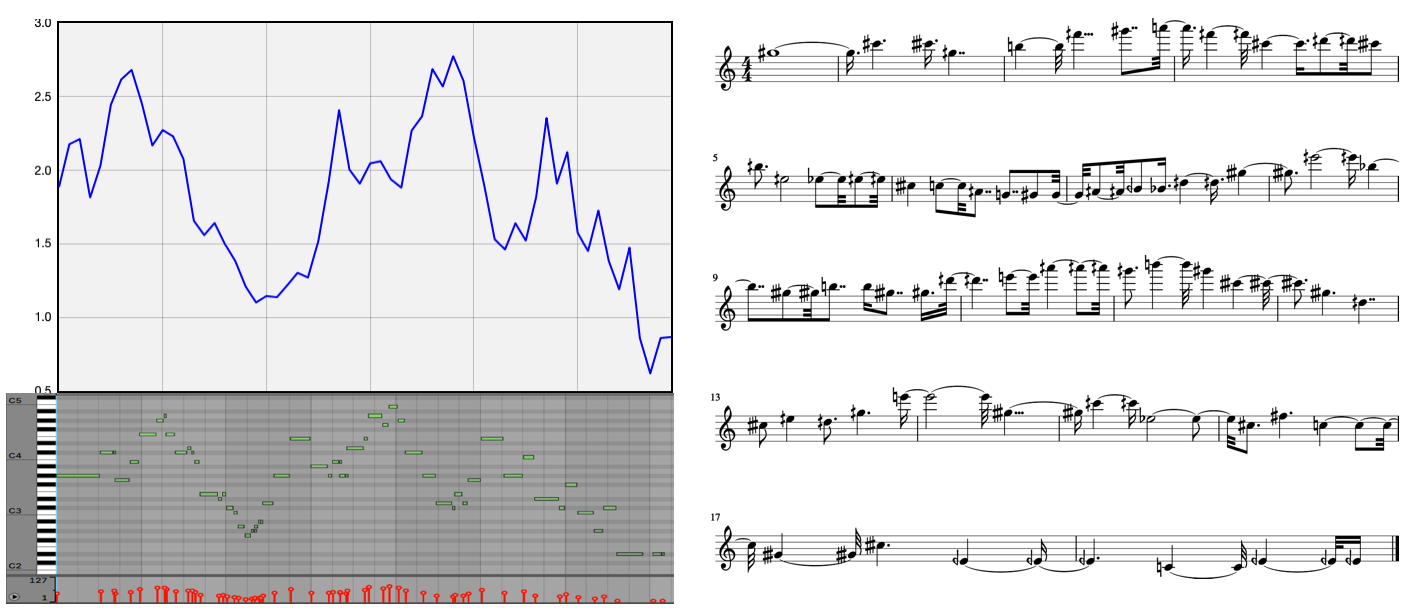}
\caption{Data, MIDI map and score from the sonification protocol in {\tt MIDIscore}} \label{sonification}
\end{figure} 

\subsection{\harmony}\label{harmony}

Utilities for harmonic analysis, design and autonomous scoring
	
See example in the notebooks on GitHub  and Ref. \cite{ref_article2} for possible uses of the modules contained here.

Harmonic analysis functions:

\begin{itemize}
\item
    {\tt def scoreFilter(seq,chords,thr=0,plot=False)}\\
	filter out low recurring chords to facilitate change point detection and score partitioning in regions. Needs the output of {\tt readScore}
	\begin{itemize}
        {\item {{\tt seq (int)}\ }} list of pcs for each chords extracted from the score
        {\item {{\tt chords (music21 {\rm object})}\ }} chords as extracted by {\tt music21}
    \end{itemize}
	{\it Returns}
	\begin{itemize}
        {\item {{\tt value, valuef (int)}\ }} integer identifier for the full and filtered chord sequence
        {\item {{\tt filtered (int)}\ }} sequence of filtered chords (by identifier as defined above)
        {\item {{\tt fmeasure (int)}\ }} measure number to which the filtered chord belongs
    \end{itemize}
\item
    {\tt def changePoint(value, vmodel='rbf', penalty=1.0, vbrakepts=None, plot = False)}\\
	run a change point detection algorithm to isolate sections on the filtered score. Uses the implementation in {\tt ruptures}.
	\begin{itemize}
        {\item {{\tt value (int)}\ }} list of pcs for each chords as extracted from  {\tt scoreFilter}
        {\item {{\tt other variables}\ }}documentation for the other variables at https://pypi.org\-/project/ruptures/
    \end{itemize}
	{\it Returns}
	\begin{itemize}
        {\item {{\tt sections (list of int)}\ }} list of breakpoints in the score sectioning (by position in the filtered chord sequence)
    \end{itemize}
\item
    {\tt def keySections(sections, GxsecDi, dnodes)} \\
	region key identification
	\begin{itemize}
        {\item {{\tt sections (int)}\ }} list of region identified by the {\\changePoint} algorithm
        {\item {{\tt GxsecDi (networkx graph {\rm object})}\ }} score sub-network as computed by {\tt scoreSubNetwork}
        {\item {{\tt dnodes ({\tt pandas} dataframe)}\ }} dataframe of node label of the full score network
    \end{itemize}
	{\it Returns}
	\begin{itemize}
        {\item {{\tt key (string)}\ }} key associated with given chord in the sequence as identified by a "prevalent chord" algorithm
        {\item {{\tt keySections (pandas dataframe)}\ }} dataframe that summarizes the key analysis results by 'section','chord range','prevalent\_chord','region'
    \end{itemize}

\item
    {\tt def tonalAnalysis(chords, sections, key, enharm=[['C','C']], write=None)}\\
	roman numeral analysis of regions in the full score
	\begin{itemize}
        {\item {{\tt sections (int)}\ }} see above
        {\item {{\tt chords (music21 {\rm object})}\ }} see above
        {\item {{\tt key (string)}\ }} see above
        {\item {{\tt enharm (list of strings)}\ }} optional enharmonic table for pitch respelling
    \end{itemize}
	{\it Returns}
	\begin{itemize}
        {\item {{\tt analysis (pandas dataframe)}\ }} dataframe that summarizes the full analysis of the score, including roman numerals - the results can be visualized 
        in musicxml format if {\tt write = True}.
    \end{itemize}

\end{itemize}

Harmonic design functions and autonomous scoring

\begin{itemize}
\item
    {\tt def chinese\_postman(graph, starting\_node)}\\
	solve the optimal routing problem of the Chinese postman on an assigned network
	\begin{itemize}
        {\item {{\tt graph (networkx graph {\rm object})}\ }} graph on which to find the optimal path
        {\item {{\tt starting\_node (int)}\ }} node from where to start the path
    \end{itemize}
	{\it Returns}
	\begin{itemize}
        {\item {{\tt graph (networkx graph {\rm object})}\ }} directed graph with optimal path (used in {\tt harmonicDesign}
    \end{itemize}

\item
    {\tt def networkHarmonyGen(mk, descriptor=None, dictionary=None, thup= None,
    thdw=None, names=None, distance=None, probs=None, write= None, pcslabel=None)} \\
	probabilistic chord distribution based on geometric distances. It is a wrapper around the modules in {\tt networks}. See the discussion in Sec. \ref{networks} for the description of the variables. Only additions are:
	\begin{itemize}
        {\item {{\tt names (list of strings)}\ }} list of operators to slice the network of pitches using {\tt vLeadNetworkByName}
        {\item {{\tt probs (list of floats)}\ }} list of probabilities to slice the network of pitches using {\tt vLeadNetwork} by distance
	\end{itemize}
	{\it Returns}
	\begin{itemize}
        {\item {{\tt nodes, edges (pandas dataframe)}\ }} nodes and edges of the generated network
    \end{itemize}
    		
\item
    {\tt def harmonicDesign(mk, nnodes, nedges, refnodes, refedges, nstart = None,
    seed=None, reverse=None, display=None, write= None)} \\
	generate a scale-free network according to the Barabasi-Albert model of preferential attachment and assign chords to nodes using the output of {\tt networkHarmonyGen}
	\begin{itemize}
        {\item {{\tt mk (musicntwrk class)}\ }} 
        {\item {{\tt nnodes, nedges (int)}\ }} number of nodes and edges per node used to generate the scale free network using {\tt networkx.barabasi\_albert\_graph}. 
        See documentation on {\tt networkx} for details.
        {\item {{\tt nodes, edges (pandas dataframe)}\ }} nodes and edges of the network generated by {\tt networkHarmonyGen}
	\end{itemize}
	{\it Returns}
	\begin{itemize}
        {\item {{\tt pitches (list of lists of int)}\ }} chord sequence of the generated network. If {\tt display = True} it draws the graph, if {\tt write = True} it writes the sequence as musicxml.
    \end{itemize}	

\item
    {\tt def rhythmicDesign(dictionary, nnodes, nedges, refnodes, refedges,
    nstart = None, seed=None, reverse=None, random=None)}\\
	builds a scale-free network according to the Barabasi-Albert model of preferential attachment and assign rhythmic figures to nodes. Details as above using the result of any of the rhythm network generation function of Sec. \ref{networks}.
	\begin{itemize}
        {\item {{\tt durations (list of lists of int)}\ }} duration sequence of the generated network. 
    \end{itemize}

\item
    {\tt def scoreDesign(pitches, durations, fac=1, TET=12, write=False)}\\
	using the sequences of pitches and rhythms from the above, generate a score in musicxml
\end{itemize}

Helper functions

\begin{itemize}
\item
    {\tt def tonalHarmonyCalculator()}\\
	multi-function harmony calculator: pitches, roman numerals, voice leading operations and more. See the in app HELP for a description of usage.

\item
    {\tt def tonalHarmonyModel(mode='minimal')} \\
	build a minimal {\it harmony model} based on voice leading operators (to be used in the calculator)

\item
    {\tt def tonnentz(x,y)}\\
	build the {\it tonnentz} for the specified x,y relation
\end{itemize}

\subsection{Plotting and general utility functions}

There are many utility functions (in {\tt utils}) that are used by other modules and that should be transparent to the average user. We recall here only the ones that are used for computing distances in pitch space, since they relate directly to the discussion of Sec. \ref{background}:

\begin{itemize}
\item
	{\tt def minimalNoBijDistance(a, b, TET, distance)}\\
    calculates the minimal distance between two pcs of different cardinality (non bijective)
    \begin{itemize}
    {\item {{\tt a,b (int)}\ }} pcs as lists or numpy arrays
    {\item {{\tt distance (str)}\ }} choice of norm in the musical space
    \end{itemize}
    	{\it Returns}
	\begin{itemize}
		{\item {{\tt dist (float)}\ }} minimal distance
        {\item {{\tt r (array of int)}\ }} multiset corresponding to minimal distance
    \end{itemize} 
\item
	{\tt generalizedOpsName(a,b,TET,distance)}\\
	finds the voice leading operator that connects two pcs (also for non bijective transformations)
	\begin{itemize}
    {\item {{\tt a,b (int)}\ }} pcs as lists or numpy arrays
    {\item {{\tt distance (str)}\ }} choice of norm in the musical space
    \end{itemize}
    	{\it Returns}
	\begin{itemize}
        {\item {{\tt r (array of int)}\ }} multiset corresponding to minimal distance
        {\item {{\tt Op (string)}\ }} VL operator that connects the two pcs
    \end{itemize} 
\end{itemize}

Finally, it is worth mentioning the network plotting utility {\tt drawNetwork} in {\tt plotting}:

\begin{itemize}
\item
	{\tt drawNetwork(nodes, edges, forceiter=100, grphtype='undirected', dx = 10, dy=10, colormap='jet', scale=1.0, drawlabels=True, giant=False)}\\
    draws the network using {\tt networkx} and {\tt matplotlib}
    \begin{itemize}
        {\item {{\tt nodes, edges (pandas dataframe)}\ }} nodes and edges of the network
        {\item {{\tt forceiter (floats)}\ }} iterations in the {\tt networks} force layout
        {\item {{\tt grphtype (string)}\ }} 'directed' or 'undirected'
        {\item {{\tt dx, dy (floats)}\ }} dimensions of the canvas
        {\item {{\tt colormap (string)}\ }} colormap for {\tt plt.get\_cmap(colormap)}
        {\item {{\tt scale (float)}\ }} scale factor for node radius
        {\item {{\tt drawlabels (logical)}\ }} draw labels on nodes
        {\item {{\tt giant (logical)}\ }} if {\tt True} draws only the giant component of the network
    \end{itemize}
\end{itemize}

The most computationally intensive parts of the modules can be run (optionally) on parallel processors using the MPI (Message Passing Interface) protocol. Communications are handled by two additional modules: {\tt communications} and\\ {\tt load\_balancing}. Since the user will never have to interact with these modules, we omit here a detailed description of their functions.

\section{Conclusions and acknowledgments}
We have presented the API for the \MUSICNTWRK\ software package. The software is freely available under GPL 3.0 and can be downloaded from GitHub and at www.musicntwrk.com, or installed directly via {\tt pip install musicntwrk}. We acknowledge the support of Aix-Marseille University, IM\'eRA, and of Labex RFIEA+. 
It must be understood that \MUSICNTWRK\ is a continuously evolving library, so it is likely that at the time of publication of this paper more functionalities will be available. We invite the reader to explore the GitHub distribution that will always provide the most recent version of the software. 
Finally, we thank Richard Kronland-Martinet, S\o lvi Ystad, Mitsuko Aramaki, Jon Nelson, Joseph Klein, Scot Gresham-Lancaster, David Bard-Schwarz, Roger Malina and Alexander Veremyer for useful discussions.

\end{document}